\documentclass{ws-procs9x6-cpt16}
\begin{document}

\newcommand{\refeq}[1]{(\ref{#1})}
\def\etal {{\it et al.}}

\title{Strongly Enhanced Effects of Lorentz-Symmetry Violation \\
in Yb$^+$ and Highly Charged Ions}

\author{M.S. Safronova}

\address{Department of Physics and Astronomy, University of Delaware\\
Newark, DE 19716, USA} 

\address{Joint Quantum Institute, NIST and the University of Maryland\\
College Park, MD 20742, USA}

\begin{abstract}
A Lorentz-symmetry test with Ca$^+$
ions demonstrated the potential of using quantum information inspired technology for tests of fundamental physics.
A systematic study of atomic-system sensitivities to Lorentz violation identified the ytterbium ion as an ideal system with high sensitivity as well as excellent experimental controllability.
A test of Lorentz-violating physics in the electron-photon sector with Yb$^+$ ions has the potential to reach levels of 10$^{-23}$, five orders of magnitude more sensitive than the current best bounds. Similar sensitivities may be also reached with highly charged ions.
\end{abstract}

\bodymatter

\section{Introduction}
Local Lorentz invariance (LLI) is an important foundation of modern physics and
has been a subject of many stringent experimental and observational tests.\cite{KosRus11} However, a number of theories unifying gravity with the Standard Model of particle physics suggest possible violation of Lorentz symmetry. While the suggested LLI-violation energy scale is much larger than the energy currently attainable by particle accelerators, it might be accessible with precision measurements at low energy. Therefore, high-precision tests of LLI with photons and particles (protons, neutrons, electrons) may provide insight into possible new physics and set limits on various theories.

  Experimental breakthroughs in atomic, molecular and optical (AMO) physics, including laser cooling and trapping of atoms, attainment of Bose-Einstein condensation, optical frequency combs, and quantum control
   led to extraordinary advances in the control of matter and light.
  These achievements, coupled with dramatic improvements in precision time and frequency metrology, measurement techniques such as atomic interferometry and magnetometry, and advances in first-principles atomic and molecular theory enabled
   a plethora of new applications of AMO, including novel ways to test the fundamental physics laws. The availability of trapped ultracold atoms and ions,
  subject to precise interrogation and control, provided for new opportunities for tests of Lorentz symmetry.
A diverse set of  AMO Lorentz-symmetry tests involves experiments with atomic clocks,\cite{WolChaBiz06} other precision spectroscopy measurements,\cite{HohLeeBud13} magnetometers,\cite{SmiBroChe11,AllHeiKar14} electromagnetic  cavities,\cite{EisNevSch09,new} and quantum information trapped ion technologies.\cite{PruRamPor15}
A cold Cs atom clock test of Lorentz invariance in the matter sector was carried out in Ref.\ \refcite{WolChaBiz06}, setting the best limits on the tensor Lorentz-violating coefficients for the proton.

 In 2015, an experiment with a pair of trapped calcium ions
improved bounds on
 LLI-violating  Standard-Model Extension (SME) coefficients $c_{JK}$ for electrons  by a factor of 100 demonstrating the potential of quantum information techniques in
the search for physics beyond the Standard Model.\cite{PruRamPor15}  The same experiment can be interpreted as testing
anisotropy in the speed of light, improving a previous such bound\cite{EisNevSch09} by a factor of 5, with the
sensitivity similar to more recent work reported in Ref.\ \refcite{new}.

\section{Ca$^+$ experiment}

Lorentz-violation tests are analyzed in the context of the phenomenological framework known as the SME, which is an effective field theory that augments the Standard-Model lagrangian with every possible combination of the Standard-Model fields that is not term-by-term Lorentz invariant, while maintaining gauge invariance, energy-momentum conservation, and observer Lorentz invariance of the total action.\cite{ColKos98}
Violations of Lorentz invariance and the Einstein equivalence principle  in bound electronic states
result in a small shift of the hamiltonian that can be described by\cite{KosLan99,HohLeeBud13}
\begin{equation}
\delta H=-\left(  C_{0}^{(0)}-\frac{2U}{3c^{2}}c_{00}\right)
\frac{\mathbf{p}^{2}}{2}-\frac{1}{6}C_{0}^{(2)}T^{(2)}_{0},\label{eq1}%
\end{equation}
where we use atomic units, $\mathbf{p}$ is the momentum of a bound electron, and $c$ is the speed of
light. The parameters $C_0^{(0)}$, $c_{00}$, and $C_{0}^{(2)}$
 are elements in the $c_{\mu \nu}$ tensor
which characterises hypothetical Lorentz violation in the electron sector within the  SME.\cite{KosLan99,KosRus11} The nonrelativistic form of the $T^{(2)}_{0}$ operator is
$T^{(2)}_{0}=\mathbf{p}^{2}-3p_{z}^{2}$.
Predicting the energy shift due to LLI violation involves the calculation of the expectation value
 of the above hamiltonian for the atomic states of interest.
Therefore, the shift of the Ca$^+$ $3d_{5/2}$ energy level due to the $c_{\mu \nu}$ tensor depends on the values of
$\langle 3d_{5/2}|\mathbf{p}^{2}| 3d_{5/2}\rangle$ and $\langle 3d_{5/2}|T^{(2)}_0|3d_{5/2} \rangle$ matrix elements.

The frequency difference (in Hz) between the shifts of the $m=5/2$ and $m=1/2$ substates of the $3d_{5/2}$ manifold was calculated in Ref.\ \refcite{PruRamPor15}:
\begin{equation}
\frac{1}{h}\left( E_{m=5/2} -E_{m=1/2}\right) = \left(-4.45(9)\times10^{15}~ \textrm{Hz} \right)\times \,C_{0}^{(2)}.
\end{equation}
 The basic idea of the Ca$^+$ experiment is to monitor this energy difference between the magnetic $m_J=|1/2|$ and $m_J=|5/2|$
 substates of the $3d_{5/2}$ $m_J$ manifold
over time to set the limit on potential
 violation of LLI.
 Therefore, the shift of the Ca$^+$ $3d_{5/2}$ energy level due to the $c_{\mu \nu}$ tensor depended only on the value  of $\langle 3d_{5/2}|T^{(2)}_0|3d_{5/2} \rangle$ matrix element, as the contribution of the scalar term canceled for the states of the same $m_J$ manifold.
 Superposition of two ions prepared in a decoherence-free subspace
 \begin{equation}
\left|\Psi \right\rangle= \frac{1}{\sqrt{2}}(\left|1/2, -1/2\right\rangle + \left|5/2, -5/2\right\rangle),
 \end{equation}
where $\left|m_1, m_2\right\rangle$ represents the $3d_{5/2}$ state with $m_J=m_1$ and $m_2$ for the first and second ion, respectively, allowed elimination of
the largest potential systematic uncertainty cased by the
fluctuation of the magnetic field.  Details of the experiment are given in Ref.\ \refcite{PruRamPor15}.

\section{Proposal for LLI test with Yb$^+$ ions}
Further improvement of LLI violation limits calls for a system with a long-lived (or ground) state that has a large $\langle j|T^{(2)}_0|j \rangle$ matrix element.
We have carried out a systematic study of this quantity for various systems and
identified general rules for the enhancement of the reduced matrix elements of the $T^{(2)}$ operator.\cite{DzuFlaSaf16}
Our calculations for  $nd$ states in Ba$^+$ and Yb$^+$, which are heavier analogues of Ca$^+$, found only a small increase of the $T^{(2)}$ matrix elements  in comparison with the Ca$^+$ case. However, Yb$^+$ has another metastable level, $4f^{13} 6s^2$~$^2F_{7/2}$ with the $T^{(2)}_0$ matrix element that is over an order of magnitude larger than for the $nd$ states.
 We find that deeper localization of the probe electron  leads to enhanced sensitivity  to the tensor Lorentz violation.
 Our study has shown that  $\langle \psi|r|\psi\rangle$ of $\sim$0.8~a.u or below for the corresponding electron is a  good indicator of  the large value of the reduced $T^{(2)}$ matrix element. This condition  is satisfied for the $4f$ hole states, such as Yb$^+$ state considered here, or for highly charged ions with $nf$ valence electrons and degree of ionization $\sim 15$.  The reduced $T^{(2)}$ matrix elements in Yb$^+$ and Sm$^{15+}$ are
135~a.u. and 149 a.u., respectively, in comparison of 9.3~a.u for the $3d_{5/2}$ state in Ca$^+$.

The Yb$^+$ $4f^{13} 6s^2$~$^2F_{7/2}$  state also has an exceptionally long lifetime on the order of several years.\cite{Huntemann2012} Therefore,  the Ramsey duration of the proposed experiment with Yb$^+$ is not limited by spontaneous decay during the measurement as in Ca$^+$ case.
The electric-octupole E3 transition between  the $4f^{13} 6s^2$~$^2\text{F}_{7/2}$ excited state and the ground state  is used as the basis for the optical atomic clock with the single trapped Yb$^+$ ion, which presently has the lowest uncertainty among the all of the optical ion clocks.\cite{Ybclock}
 Yb$^+$ ions are also used in quantum information research.\cite{IslCamKor13} As a result, experimental techniques for precision control and manipulation of Yb$^+$ atomic states are particulary well developed making it an excellent candidate for searches of Lorentz-violation signature.

We estimated that experiments with the metastable $4f^{13} 6s^2$~$^2\text{F}_{7/2}$  state of Yb$^+$  can reach sensitivities of $1.5\times10^{-23}$ for the $c_{JK}$ coefficients,\cite{DzuFlaSaf16} over $10^5$ times more stringent than current best limits.\cite{PruRamPor15} Moreover, the projected sensitivity to the $c_{TJ}$ coefficients will be at the level of
$1.5\times 10^{-19}$, below the ratio between the electroweak and Planck energy scales.\cite{DzuFlaSaf16}
Similar sensitivities may potentially be reached for LV tests with highly charged ions, given future development of experimental techniques
for these systems.

\end{document}